paper

 \documentstyle[preprint,eqsecnum,aps,amsfonts]{revtex}
 \tighten
\begin{document}
\title{Dimensional perturbation theory for vibration-rotation
spectra\\ of linear triatomic molecules}
\author{Andrei A. Suvernev and David Z. Goodson}
\address{Department of Chemistry, Southern Methodist University,
Dallas, Texas  75275}
\date{\today}
\maketitle
\begin{abstract}
A very efficient large-order perturbation theory
is formulated for the nuclear motion of a linear triatomic
molecule.  To demonstrate the method,
all of the experimentally observed rotational energies,
with values of $J$ almost up to 100,
for the ground and first excited vibrational states of
CO$_2$ and for the ground vibrational states
of N$_2$O and of OCS are calculated.  All coupling between
vibration and rotation is included.  The perturbation
expansions reported here are rapidly convergent.  
The perturbation parameter is $D^{-1/2}$, where $D$ is
the dimensionality of space.  Increasing $D$ is qualitatively
similar to increasing the angular momentum quantum number $J$.
Therefore, this approach is especially suited for states
with high rotational excitation.  The computational cost of
the method scales only as $JN_v^{5/3}$, where
$N_v$ is the size of the vibrational basis set.
\\
\end{abstract}

\narrowtext

\section{Introduction}

The calculation of highly excited rovibrational states
of molecules is a challenging computational problem.  The
traditional method is numerical diagonalization of the
Hamiltonian in a finite basis set.  This works well for low-lying
states, but for highly excited states it is tractable only with
judicious choice of basis set \cite{wolfsberg,ten} and the
use of special numerical techniques \cite{light,tenhen,tennyson},
on account of the extremely large matrices that are needed in
order to obtain sufficient accuracy.

It has been suggested \cite{cizek} that large-order
perturbation theory is a viable
alternative approach to calculating vibrational spectra.
We recently compared numerical diagonalization with large-order
perturbation theory in powers of the coupling constant,
for a model system of coupled anharmonic oscillators, and
found that the perturbation theory had a significantly lower
computational cost \cite{suvernev}.  Here we apply {\it dimensional}
perturbation theory (DPT) to the calculation of vibration-rotation
spectra for actual triatomic molecules.  This is a type of
perturbation theory that is especially suited for
states with high angular momentum \cite{kais}.

This theory \cite{book1,book2,yaffe,chatterjee,drhrev}
is one example of a class of semiclassical perturbation
expansions that have been applied to problems in high-energy
physics, statistical mechanics,  solid state physics, and
polymer science.  There has been much recent interest in using
DPT to solve the electronic Schr\"odinger equation for atoms.  
At the zeroth order of the perturbation theory an atom becomes
a rigid rotor, with the electrons stationary relative to each
other.  At first order the electrons vibrate harmonically.
The first-order results for electronic energies of atoms are
typically accurate to within 1\% \cite{loeser}, and
computational techniques have now been developed that make it
possible to carry the perturbation theory to much higher order
\cite{vainberg,dunn,gscf}.

Thus far there have been few attempts to apply this theory
to molecules, despite the resemblance between DPT and the
standard qualitative picture of molecular vibrations.  First-order
expansions have been calculated for the electronic energy
of H$_2$ \cite{frantz}, for the full electronic-vibrational
problem (without the Born-Oppenheimer approximation) for
H$_2^{\,+}$ \cite{cohen}, and for the vibrational motion
of atomic clusters \cite{clusters}.  These studies have
provided intriguing qualitative models of correlated
many-particle dynamics and semiquantitative results for
energy eigenvalues.  Recently, however,
Kais and Herschbach \cite{kais} used
DPT to compute the rotation spectrum of the ground
vibrational state of H$_2$.  The
accuracy of their results, from just first-order perturbation
theory, remained excellent even for $J$ as
high as 50.

The distinctive feature of DPT is that it uses a
Hamiltonian operator that has been analytically continued to
a nonphysical hyperdimensional coordinate space.  
Typically, the kinetic energy operator is generalized to
a $D$-dimensional form while the potential energy is
left unchanged.  Since this
dimensional continuation is nonphysical, it allows for a great
deal of flexibility in how one chooses to define it.
We will formulate the perturbation theory in such a way
that it closely resembles the Kais-Herschbach diatomic analysis.
In Section~\ref{sec:ke} we develop our dimensional
continuation of the kinetic energy operator.
Section~\ref{sec:pt} describes the method we have used to
compute the expansion coefficients.  In Section~\ref{sec:results}
we use the theory to reproduce the empirical rotational spectra
for various linear triatomic molecules and in
Section~\ref{sec:conclusions} we discuss advantages
of this approach.

\section{Dimensional continuation of the Kinetic Energy Operator}
\label{sec:ke}

DPT is based on the fact that the dynamics of a system of
particles simplifies as the dimensionality becomes
very large.  This simplication depends on the partitioning of 
the coordinates into translational, rotational, and internal
degrees of freedom.  Consider an arbitrary system of $N$
particles in $D$ spatial dimensions and let $\tau_{\rm trans}$,
$\tau_{\rm rot}$, and $\tau_{\rm int}$ be the number of
coordinates of the indicated type.  $\tau_{\rm trans}$ and
$\tau_{\rm rot}$ increase linearly with $D$ while
$\tau_{\rm int}$ quickly reaches a maximum value of $N(N-1)/2$
and then remains constant with increasing $D$.  Since the
potential energy depends only on the internal coordinates,
the energy levels of the system can in principle be determined
from a differential equation in just $\tau_{\rm int}$
coordinates even in the limit $D\to\infty$.

The separation of the $D$ translational coordinates is
straightforward.  Consider a system of 3 atoms, which we will
describe in terms of the center-of-mass vector,
${\bf r}_{\rm cm}$; the vector between atoms 1 and 2, 
${\bf r}_{\rm s}$; and the vector ${\bf r}_{\rm a}$ that leads
from the center of mass of atoms 1 and 2 to particle 3.
If ${\bf r}_1$, ${\bf r}_2$ and ${\bf r}_3$ specify the
positions of the atoms relative to the center of mass, the
${\bf r}_{\rm s}={\bf r}_2-{\bf r}_1$ and
\begin{equation}
{\bf r}_{\rm a}
={\bf r}_3-[\,(m_1/m_{\rm s})\,{\bf r}_1+(m_2/m_{\rm s})
\,{\bf r}_2\,],
\end{equation}
where $m_{\rm s}=m_1+m_2$.  The the kinetic energy operator
can be written as
\begin{equation}
T=-{\scriptstyle{1\over 2}}\Delta_{{\bf R}_{\,\rm cm}}+T_{\rm rel},
\quad
T_{\rm rel}=-{\scriptstyle{1\over 2}}
\left(\Delta_{{\bf R}_{\,\rm s}}+\Delta_{{\bf R}_{\,\rm a}}\right),
\label{T}
\end{equation}
in terms of the mass-normalized coordinates
\begin{mathletters}
\begin{eqnarray}
&&{\bf R}_{\,\rm cm}=m^{1/2}\,{\bf r}_{\rm cm},\\
&&{\bf R}_{\,\rm s}=(m_1m_2/m_{\rm s})^{1/2}\,{\bf r}_{\rm s},\\
&&{\bf R}_{\,\rm a}=(m_{\rm s}m_3/m)^{1/2}\,{\bf r}_{\rm a},
\end{eqnarray}
\label{Rcoords}
\end{mathletters}

\noindent
where $m=m_1+m_2+m_3$ is the total mass of the system.

The separation of rotational coordinates is more complicated,
but is key to the simplying effect of increasing $D$.  The
qualitative effect of separating out
angular coordinates is to introduce a centrifugal potential
into the internal-coordinate kinetic energy operator.
The linear increase with $D$ in the number of angular degrees
of freedom causes the strength of the centrifugal potential
to increase quadratically with $D$.  The derivatives with
respect to the internal coordinates are overwhelmed by the
centrifugal potential in the limit $D\to\infty$.  This
elimination of the derivatives yields a Schr\"odinger equation
that can be solved exactly even for many-particle systems.

Consider first a {\it diatomic} molecule with mass-normalized
internuclear distance ${\bf R}_{\,\rm s} $.  If we define
the $D$-dimensional Laplacian as
\begin{equation}
\Delta_{{\bf R}_{\,\rm s}}
=\sum_{i=1,D}{\partial^2\over\partial x_i^2},
\label{laplacian}
\end{equation}
where the $x_i$ are the Cartesian components of
${\bf R}_{\,\rm s}$, then the Schr\"odinger equation can be
expressed as \cite{louck,tutorial}
\begin{equation}
\left[ -{1\over 2}\Delta_{{\bf R}_{\,\rm s}}
+V(R_{\rm s})-E\right]
\Phi(R_{\rm s})Y_{J,M,D-1}({\bf\Omega}) = 0,
\label{diatomicscheq0}
\end{equation}
where
\begin{equation}
\Delta_{{\bf R}_{\,\rm s}}
={1\over R_{\rm s}^{D-1}}{d\over dR_{\rm s}}
 \left( R_{\rm s}^{D-1}{d\over dR_{\rm s}}\right)
-{J(J+D-2)\over R_{\rm s}^2}.
\label{diatomicdelta}
\end{equation}
The $Y_{J,M,D-1}$, which are hyperdimensional generalizations of
the spherical harmonics, are functions of the set $\bf\Omega$ of
$D-1$ independent angular coordinates.  The substitution
\begin{equation}
\Phi(R_{\rm s})=R_{\rm s}^{-(D-1)/2}\Psi(R_{\rm s})
\label{diatomicpsi}
\end{equation}
removes the first derivatives, with the result
\begin{equation}
\left[ -{1\over 2}{d^2\over dR_{\rm s}^2}
 + {J_D(J_D+1)\over 2R_{\rm s}^2} +V-E\right]\Psi = 0,
\label{diatomicscheq}
\end{equation}
where
\begin{equation}
J_D=J+(D-3)/2.
\label{jd}
\end{equation}
This is the standard approach to dimensional continuation
\cite{herrick,mlodinow} in which the kinetic energy is defined
in a $D$-dimensional Cartesian coordinate system while the
potential energy is left in its 3-dimensional form (except,
perhaps, for certain scale factors
\cite{mlodinow,watson,dscaling}).
It is straightforward to use Eq.~(\ref{diatomicscheq}) to
calculate a large-order asymptotic expansion for $E$ in powers
of $1/D$ or $1/J_D$ \cite{vainberg,dunn}.

For triatomics this analysis is complicated by the fact that
the total angular momentum of the system depends not just on
the external rotation coordinates but also on the angle
between ${\bf R}_{\,\rm s}$ and ${\bf R}_{\,\rm a}$.  Rather
than carry out the separation of variables with the
$D$-dimensional Laplacian, we will use a simpler approach.
There are only two considerations that restrict the way in
which the dimensional continuation is defined: The
Schr\"odinger equation must be the correct physical equation
for $D=3$, and the large-$D$ limit must be a reasonable, and
solvable, zeroth-order approximation for the perturbation
theory.  Our strategy is to perform the separation of
variables using the 3-dimensional Laplacian and then introduce
an arbitrary $D$ dependence consistent with these considerations.
In essence, we will allow one body-fixed axis to rotate in
$D$ dimensions while a second body-fixed axis rotates about
the first in a 3-dimensional space.

For the internal coordinates we will use $R_{\rm s}$ and the
cylindrical coordinates $\rho$ and $z$,
\begin{equation}
\rho^2+z^2=R_{\rm a}^2,\quad \rho/z=\tan \beta
\label{cylindrical}
\end{equation}
where $\beta$ is the angle between ${\bf R}_{\,\rm a}$
and ${\bf R}_{\,\rm s}$.
The derivation of the 3-dimensional kinetic energy operator in
these coordinates is given in Appendix \ref{app:kineticenergy}.
Let $\hat{L}$ be the total angular
momentum operator, $\hat{L}_1$ the operator for
the projection of angular momentum onto a space-fixed axis
${\bf x}_1$, and $\hat{K}$
the operator for the projection onto the body-fixed axis
${\bf R}_{\,\rm s}$, and let $J$, $M$, and $K$ be the
corresponding quantum numbers.  It is convenient to describe
the wavefunction in terms of symmetrized basis functions
\begin{equation}
\phi^{(\pm )}_{J,M,K}
=2^{-1/2}\left( \phi_{J,M,K}\pm\phi_{J,M,-K}\right),
\label{basisfns}
\end{equation}
where the $\phi_{J,M,K}(R_{\rm s},\rho,z)$ are simultaneous
eigenfunctions of $\hat{L}$, $\hat{L}_{x_1}$, and $\hat{L}_{\rm s}$,
and $K$ is now defined to be nonnegative.
Let $\sigma$ be a parity index that indicates the sign in
Eq.~(\ref{basisfns}).  Then the internal-coordinate part
of the kinetic energy operator for given $J$ and $M$ can be
expressed in terms of the matrix elements
${\cal T}_{J;K,\sigma;K'\! ,\sigma'}
=\langle K,\sigma|T_{\rm rel}|K'\! ,\sigma'\rangle$.
In Appendix \ref{app:kineticenergy} we derive the following
expression:
\widetext
\begin{eqnarray}
{\cal T}_{J,K,\sigma}
=\delta_{\sigma'\!,\sigma}\delta_{K'\!,K}
&&\left[T^{(1)}_{J,K}(R_{\rm s\!})+T^{(2)}_K(\rho,z)
+{1\over 2R_{\rm s}^2}T^{(3)}_K(\rho,z)\right]
\nonumber\\
&&\qquad\qquad\qquad\qquad\quad
+\;\delta_{\sigma',-\sigma}\,{1\over 2R_{\rm s}^2}
\Big[ \delta_{K'\! ,K+1}\, T^{(4,+)}_{J,K,\sigma}(\rho,z)
+\delta_{K'\! ,K-1}\, T^{(4,-)}_{J,K,\sigma}(\rho,z)\Big] ,
\label{relke}
\end{eqnarray}

\narrowtext
\noindent
where
\begin{mathletters}
\begin{eqnarray}
&&T^{(1)}_{J,K}=
-\frac{1}{2R_{\rm s}^2}
\frac{\partial}{\partial R_{\rm s}}R_{\rm s}^2
\frac{\partial}{\partial R_{\rm s}} 
+{J(J+1)-K^2\! -1\over 2R_{\rm s}^2},\! \\
\label{t1}
&&T^{(2)}_K=
- {1\over 2}\frac{\partial^2}{\partial z^2}
- \frac{1}{2\rho}\frac{\partial}{\partial\rho}
   \rho\frac{\partial}{\partial\rho}+\frac{K^2}{2\rho^2},\\
&& T^{(3)}_K=
-\rho^2\frac{\partial^2}{\partial z^2}
-z^2\left(\frac{1}{\rho}\frac{\partial}{\partial\rho}
 \rho\frac{\partial}{\partial\rho}
-{K^2\over\rho^2}\right)
\nonumber\\
&&\qquad\qquad\qquad\qquad
+\left( 2z \frac{\partial}{\partial z}+1\right)
 \left(\rho\frac{\partial}{\partial \rho}+1\right),\\
&& T^{(4,+)}_{J,0,\sigma} = -\sqrt{2J(J+1)}
\left( \rho{\partial\over\partial z}
 -z{\partial\over\partial\rho}-{z\over 2\rho}\right),\\
&& T^{(4,-)}_{J,0,\sigma} = 0,\quad
T^{(4,-)}_{J,1,-} = 0,\\
&& T^{(4,-)}_{J,1,+} = \sqrt{2J(J+1)}
\left( \rho{\partial\over\partial z}
 -z{\partial\over\partial\rho}\right),\\
&& T^{(4,\pm )}_{J,K,\sigma} =
\mp\sqrt{J(J+1)-K(K \pm 1)}
\nonumber\\
&&\qquad\qquad\qquad\quad \times
 \bigg[\rho\frac{\partial}{\partial z}
   -z\frac{\partial}{\partial \rho}
    \mp (K \pm 1)\frac{z}{\rho}\bigg] ,
\label{t4}
\end{eqnarray}
\end{mathletters}

\noindent
with Eq.~(\ref{t4}) valid for $K>1$ and for the case
$K=1$, $\sigma=+$.  Note that there will be two independent
sets of $\phi_{J,M,K}^{(\pm)}$, on account of the form of
the Kronecker deltas that multiply the
$T_{J,K,\sigma}^{(4,\pm)}$ in Eq.~(\ref{relke}).  One set
will consist of basis functions for which $\sigma=+$ and $K$ is
even or $\sigma=-$ and $K$ is odd, while the other will
consist of those functions with $\sigma=-$ and $K$ even or
$\sigma=+$ and $K$ odd.  The lowering operator
$T_{J,1,\sigma}^{(4,-)}$ is zero for $\sigma=-$ but nonzero for 
$\sigma=+$ because of the fact that $\phi_{J,M,0}^{(-)}=0$.

The only modification we will make in the 3-dimensional operator
${\cal T}_{J,K,\sigma}$ is to replace $T^{(1)}_{J,K}$ with
the dimensional continuation
\begin{eqnarray}
T^{(1)}_{J,K}&=&
-\frac{1}{2R_{\rm s}^{(D-1)}}
\frac{\partial}{\partial R_{\rm s}}
\left(R_{\rm s}^{(D-1)}\frac{\partial}{\partial R_{\rm s}}\right)
\nonumber\\
&&\qquad\qquad
+{J(J+D-2)-K^2-1\over 2R_{\rm s}^2},
\label{newt1}
\end{eqnarray}
which is analogous to the diatomic result in
Eq.~(\ref{diatomicdelta}).
If we use the scaled basis functions
\begin{equation}
\psi_{J,K,M}^{(\pm)}(R_{\rm s},\rho,z)
=R_{\rm s}^{(D-1)/2}
\phi_{J,M,K}^{(\pm)}(R_{\rm s},\rho,z),
\end{equation}
then $T^{(1)}_{J,K}$ becomes
\begin{equation}
\widetilde{T}^{(1)}_{J,K}=
-{1\over 2}{\partial^2\over\partial R^2_{\rm s}}
+{1\over 2 R_{\rm s}^2}
 \left[ J_D(J_D\! +1)-K^2\! -1)\right],\!\!\\
\label{scaledt1}
\end{equation}
with $J_D$ given by Eq.~(\ref{jd}).

\section{The Perturbation Theory}
\label{sec:pt}

\subsection{Dimensional Scaling}

We consider here systems that can be described by a potential
of the form
\begin{eqnarray}
U&=&
\frac{1}{2} \,\omega_{\rm a}^2\,(z-z_{\rm eq})^2
+\frac{1}{2}\,\omega_{\rm b}^2 \,\rho^2
\nonumber\\
&&\qquad
+\frac{1}{2} \,
\omega_{\rm s}^2\,\bigl(R_{\rm s}-R_{\rm eq}\bigr)^2
+ \lambda \,\bigl(R_{\rm s}-R_{\rm eq}\bigr)\,\rho^2,
\label{potential}
\end{eqnarray}

\noindent
which is appropriate for a linear triatomic molecule.
We will assume that $U$ is independent of dimension.

Our goal is to derive an asymptotic expansion about the limit
$D\to\infty$ in terms of the parameter
\begin{equation}
\kappa=D^{-1/2},
\label{kappa}
\end{equation}
which will be summed at the physical value $\kappa=3^{-1/2}$.
To obtain a useful $\kappa\to 0$ limit, we introduce the
dimension-scaled quantities
\begin{equation}
r_{\rm eq}=\kappa R_{\rm eq},
\quad
\tilde{\lambda}=\kappa^{-1}\lambda,
\quad
\widetilde{E}=\kappa^2 E,
\label{dscalings}
\end{equation}
and the displacement coordinates
\begin{equation}
r=\kappa (R_{\rm s}-R_{\rm eq}),
\quad y=z-z_{\rm eq}.
\end{equation}
We will treat $r_{\rm eq}$, $\tilde{\lambda}$ and
$\widetilde{E}$ as $\kappa$-independent constants with
values $3^{-1/2}R_{\rm eq}$, $3^{1/2}\lambda$,
and $3^{-1}E$, respectively.

The dimension-scaled Hamiltonian matrix can be expressed
in terms of the operator
\begin{eqnarray}
{\cal H} &=& \delta_{\sigma'\!,\sigma}\delta_{K'\!,K}
\left[ {\cal H}^{({\rm diag})}+\kappa^2\tilde{\lambda}r\rho^2
+\kappa^4{1\over 2(r_{\rm eq}+r)^2}T_K^{(3)}\right]
\nonumber\\
&&
 +\ \kappa^4 \delta_{\sigma'\!,-\sigma}
 \left( \delta_{K'\!,K+1}T^{(4,+)}_{J,K,\sigma}
 \! +\delta_{K'\!,K-1}T^{(4,-)}_{J,K,\sigma}\right) ,
\label{scaledH}
\end{eqnarray}
where ${\cal H}^{({\rm diag})}$ is the separable operator
\begin{eqnarray}
&&{\cal H}^{({\rm diag})} =
\kappa^2\, T^{(2)}_K
-\kappa^4{1\over 2}{\partial^2\over\partial r^2}
\nonumber\\
&&
\quad +\,{1\over 2}\omega_{\rm s}^2 r^2
+\kappa^2{1\over 2}\big(\omega_{\rm a}^2y^2
 +\omega_{\rm b}^2\rho^2\big)
\nonumber\\
&&
\quad +\, {1+4(J-1)\kappa^2+4(J^2\! -K^2\! -2J-1/4)\kappa^4
     \over 8(r_{\rm eq}+r)^2}.
\nonumber\\
\label{Hdiag}
\end{eqnarray}

\noindent
In the limit $\kappa\to 0$ all derivatives in
Eq.~(\ref{scaledH}) drop out, leaving the system localized
at $\rho=0$, $y=0$, and $r=r_{\rm min}$, where
$r_{\rm min}$ is the minimum of the effective potential
\hbox{$(1/8)(r_{\rm eq}+r)^{-2}+(1/2)\omega_{\rm s}^2r^2$.}  The
numerical value of $r_{\rm min}$ can be determined from the
equation
\begin{equation}
4\omega_{\rm s}^2\,r_{\rm min}\,(r_{\rm eq}+r_{\rm min})^3
-1=0.
\label{minimum}
\end{equation}

\subsection{The Energy Expansion}

The next step is to introduce a dimension-scaled
displacement coordinate $x$,
\begin{equation}
r = r_{\rm min} + \kappa x
\label{scaling}
\end{equation}
and expand Eq.~(\ref{scaledH}) in powers of $\kappa$,
\begin{equation}
{\cal H}=E_0+\kappa^2\sum_{k=0}^\infty
\kappa^k{\cal H}_k,
\label{Hexpansion}
\end{equation}
where $E_0$ is the zeroth-order solution for $\widetilde{E}$,
\begin{equation}
E_0=\raise1pt\hbox{$\scriptstyle{1\over 4}$}\tilde{B}
+\raise1pt\hbox{$\scriptstyle{1\over 2}$}
 \omega_{\rm s}r_{\rm min}^2,
\label{e0}
\end{equation}
in terms of the effective rotational constant
\begin{equation}
\tilde{B}={1\over 2(r_{\rm eq}+r_{\rm min})^2}.
\end{equation}
At order $\kappa^2$ we have a separable Schr\"odinger
equation that can be solved exactly.  Our strategy will be
to express the wavefunction at any order in $\kappa$
in the form of a linear combination of the separable product
eigenfunctions obtained at order $\kappa^2$.

Note that several of the terms in Eqs.~(\ref{scaledH}) and
(\ref{Hdiag}) are proportional to $\kappa^4$.  We can improve
the convergence of our perturbation theory by redefining the
dimensional continuation so that these terms enter at lower
orders \cite{suvernevlet}.  Consider the operator
${\cal H}^{({\rm diag})}$.  The
second derivative with respect to $r$, after the change of variable
in Eq.~(\ref{scaling}), becomes proportional to $\kappa^2$.
We will replace the other factor of $\kappa^4$ in
Eq.~(\ref{Hdiag}) with $\kappa^2/3$, taking advantage of the
fact that our Hamiltonian will still be valid as long as it
is correct for the physical case $\kappa=3^{-1/2}$.  In
Eq.~(\ref{scaledH}) we will replace the factors of $\kappa^4$
with $3^{-1/2}\kappa^3$, in order to still have a
separable solution at order $\kappa^2$.

Thus, at order $\kappa^2$, we have the operator
\begin{equation}
{\cal H}_0 = \delta_{\sigma',\sigma}\delta_{K'\! ,K}
({\cal T}_0+{\cal V}_0),
\end{equation}
where
\begin{equation}
{\cal T}_0 =
-{1\over 2}\left(
{\partial^2\over \partial x^2}+
{\partial^2\over \partial y^2} \right)
-{1\over 2}\left({1\over\rho}{\partial\over\partial\rho}
 \rho{\partial\over\partial\rho} - {K^2\over\rho^2}\right)
\end{equation}
and
\begin{eqnarray}
{\cal V}_0 &=&
+{1\over 2}\tilde{\omega}_{\rm s}^2x^2
+{1\over 2}\omega_{\rm a}^2y^2
+{1\over 2}\tilde{\omega_{\rm b}^2}\rho^2
\nonumber\\
&&\qquad\quad
+{1\over 3}\big[ J(J+1)-(K^2+13/4)\big]\, \tilde{B} ,
\label{h0}
\end{eqnarray}

\noindent
with the effective frequencies $\tilde{\omega}_{\rm s}$
and $\tilde{\omega}_{\rm b}$ given by
\begin{equation}
\tilde{\omega}_{\rm s}^2=\omega_{\rm s}^2+3\tilde{B}^2,
\qquad
\tilde{\omega}_{\rm b}^2=\omega_{\rm b}^2
+2\tilde{\lambda}r_{\rm min}.
\end{equation}
The normalized eigenfunctions of ${\cal H}_0$ are
\widetext
\begin{eqnarray}
f_{{\bf n},K}\,(x,y,\rho) &=&
\frac{(\tilde{\omega}_{\rm s}\,\omega_{\rm a})^{1/4}
(\tilde{\omega}_{\rm b}\;n_{\rm b}!)^{1/2}}
{2^{(n_{\rm s}+n_{\rm a}-1)/2}\,[\pi\,n_{\rm s}!
\,n_{\rm a}!\,(n_{\rm b}+K)!]^{1/2}}
\nonumber\\
&&\qquad \times\,
\big(\tilde{\omega}_{\rm b}^{1/2}\rho \big)^K
\;L_{n_{\rm b}}^{K}\! (\tilde{\omega}_{\rm b} \,\rho^2)
H_{n_{\rm s}}\! (\tilde{\omega}_{\rm s}^{1/2} x)
\;H_{n_{\rm a}}\! ( \omega_{\rm a}^{1/2} y)
\, e^{-(\tilde{\omega}_{\rm s} \,x^2
 +\omega_{\rm a}\,y^2+\tilde{\omega}_{\rm b} \,\rho^2)/2},
\label{eigenfns}
\end{eqnarray}

\noindent
where the $H_n$ are Hermite polynomials and the $L_n^K$
are generalized Laguerre polynomials.  ${\bf n}$ indicates
the ordered set of vibrational quantum numbers
$(n_{\rm s},n_{\rm a},n_{\rm b})$.  The eigenvalues are
\begin{equation}
\epsilon_0^{({\bf n},J,K)}=
\left( n_{\rm s}+ {1\over 2}\right)
\tilde{\omega}_{\rm s}
+(2n_{\rm b}+K+1)\,\tilde{\omega}_{\rm b}
+\left( n_{\rm a} + {1\over 2}\right)\omega_{\rm a}
+ {1\over 3}
\left[ J(J+1)-\left( K^2+{13\over 4}\right)\right] \tilde{B} .
\label{eigenvalues}
\end{equation}
\narrowtext

At this order, $K$ is a good quantum number.  Thus, we can
unambiguously label the eigenstates according to ${\bf n}$,
$J$, $M$, and $K$.  The energy levels,
\begin{equation}
E\sim 3\,(E_0+\epsilon_0/3),
\end{equation}
are labeled by ${\bf n}$, $J$, and $K$.
The operators $T^{(4,\pm)}$, which enter the analysis at
order $\kappa^3$, will cause the wavefunction
to contain contributions from basis functions
$\phi^{(\sigma)}_{J,M,K'}$ with $K'$ different from the
initial choice of $K$, but each initial choice will correspond
to a physically distinct energy level.  Note that
$\epsilon_0^{({\bf n},J,K)}$ does not depend on $\sigma$.
Higher-order terms will break this degeneracy, giving rise
to the so-called ``$l$-type doubling'' \cite{allen}.

Perturbation equations for the large-order analysis are obtained
from the Schr\"o\-dinger equation by subtituting
Eq.~(\ref{Hexpansion}) for the Hamiltonian, and then
substituting an expansion of the form
\begin{equation}
\widetilde{E}=E_0+\kappa^2\sum_{k=0}^\infty\kappa^k\epsilon_k
\end{equation}
for the energy, an expansion of the form
\begin{equation}
\Phi^{({\bf n},J,K,\sigma)}=
R_{\rm s}^{-(D-1)/2}\sum_{k=0}^\infty
\kappa^k\Psi_k^{({\bf n},J,K,\sigma)}(x,z,\rho)
\end{equation}
for the wavefunction, and then collecting terms by powers of
$\kappa$.  The zeroth-order solutions for the wavefunction are
\begin{equation}
\Psi_0^{({\bf n},J,K,\sigma )}=f_{{\bf n},K},
\end{equation}
unless $K=0$ and $\sigma=-$, in which case it is zero,
according to Eq.~(\ref{basisfns}).  The
$\epsilon_0^{({\bf n},J,K)}$ are given by Eq.~(\ref{eigenvalues}).

The higher-order results can be obtained recursively.
Let us use unnormalized wavefunctions subject to the condition
\cite{orthocondition}
\begin{equation}
\big\langle \Psi_0\big|\Psi_k\big\rangle=\delta_{k,0}.
\label{condition}
\end{equation}
Then the perturbation equations can be rearranged to obtain
a solution for $\Psi_k$ in terms of the $\Psi_{j<k}$.
For given choice of ${\bf n}$, $J$, $K$, and $\sigma$, let
\begin{equation}
\Psi_k=\sum_{{\bf n}',\sigma',K'}
a_k^{(K'\!,\sigma'\!\!,{\bf n}')} f_{{\bf n}',K'}\ .
\label{lincom}
\end{equation}
The range of the $K'$ summation is from 0 to $J$, and
if $K'\ne 0$, then $\sigma' =+$ or $-$.  
The ranges of the $n'_\alpha$ depend on the value of $k$ and on
the maximum order to which the energy expansion is to be
calculated \cite{dunn}.  For a calculation up to $\epsilon_{2p}$,
for given $p$, the wavefunction term with the largest number
of components is $\Psi_p$, for which $n'_\alpha\le 3p+n_\alpha$.
At order $\kappa^k$ we have
\widetext
\begin{eqnarray}
&&a_k^{(K'\!,\sigma'\!\!,{\bf n}'\! )} =
\ {1\over C_{K'\!,{\bf n}'}}\sum_{k'=1}^k
\Bigg[ \epsilon_{k'}a_{k'}^{(K'\! ,\sigma'\! ,{\bf n}')}
\nonumber\\
&&\qquad - \sum_{\,{\bf n}''}
\bigg( H^{(0)}_{k'\!,K'\!,{\bf n}'\!,{\bf n}''}
 a_{k-k'}^{(K'\!,\sigma'\!\!,{\bf n}'')}
 + H^{(+)}_{k'\!,K'\!\!,{\bf n}'\!,{\bf n}''}
 a_{k-k'}^{(K'\!+1,-\sigma'\!\!,{\bf n}'')}\!
+ H^{(-)}_{k'\!,K'\!\!,{\bf n}'\!,{\bf n}''}
 a_{k-k'}^{(K'\!-1,-\sigma'\!\!,{\bf n}'')}\!\bigg)\Bigg],
\label{arecursion}
\end{eqnarray}

\noindent
where
\begin{equation}
C_{K'\!,{\bf n}'}=
(n'_{\rm s}-n_{\rm s})\,\tilde{\omega}_{\rm s}
+(n'_{\rm a}-n_{\rm a})\,\omega_{\rm a}
+(2n'_{\rm b}+K'-2n_{\rm b}-K)\,\tilde{\omega}_{\rm b},
\end{equation}
\begin{mathletters}
\begin{eqnarray}
&& H^{(0)}_{k,K,{\bf n},{\bf n}'} =
(-1)^k {1\over 8}\tilde{\omega}_{\rm s}\gamma^2
\bigg\{
(k+3)\gamma^{k+2} \hat{q}^{k+2}_{n_{\rm s},n'_{\rm s}}
+{4\over 3}\Big[ J(J+1)-(K^2+13/4)\Big](k+1)\gamma^k
  \hat{q}^k_{n_{\rm s},n'_{\rm s}}
\nonumber\\
&&\qquad\qquad\qquad\quad
+\, 4k\gamma^{k-1}
  \hat{q}^{k-1}_{n_{\rm s},n'_{\rm s}}
  \bigg[
  \left({\omega_{\rm a}\over\tilde\omega_{\rm b}}\right)
  \hat{p}^2_{n_{\rm a},n'_{\rm a}}\hat{b}_{K,n_{\rm b},n'_{\rm b}}
 -\left({\tilde\omega_{\rm b}\over\omega_{\rm a}}\right)
  \hat{z}^2_{n_{\rm a},n'_{\rm a}}\hat{c}_{K,n_{\rm s},n'_{\rm s}}
 + \hat{d}_{n_{\rm a},n'_{\rm a}}
   \hat{e}_{K,n_{\rm b},n'_{\rm b}} \bigg] \bigg\}
\nonumber\\
&&\qquad\qquad\qquad\quad
\ +\,\delta_{k,1}\,
    {3\lambda
      \over\tilde{\omega}_{\rm s}^{1/2}\tilde{\omega}_{\rm b}}\,
         \hat{q}_{n_{\rm s},n'_{\rm s}}
         \hat{b}_{K,n_{\rm b},n'_{\rm b}},
\\
&& H^{(\pm)}_{k,K} =
(-1)^k {1\over 2} \tilde{\omega}_{\rm s}
\sqrt{J(J+1)-K(K\pm 1)}
\nonumber\\
&&\qquad\qquad\qquad\qquad
\times\, k \gamma^{k+1}
\hat{q}^{k-1}_{n_{\rm s},n'_{\rm s}}
 \bigg[
 \left({\omega_{\rm a}\over\tilde\omega_{\rm b}}\right)^{1/2}
    \! \hat{p}_{n_{\rm a},n'_{\rm a}}
       \hat{f}^{(\pm )}_{K,n_{\rm b},n'_{\rm b}},
 -\left({\tilde\omega_{\rm b}\over\omega_{\rm a}}\right)^{1/2}
   \! \hat{z}_{n_{\rm a},n'_{\rm a}}
      \hat{g}^{(\pm )}_{K,n_{\rm b},n'_{\rm b}} \bigg],
\end{eqnarray}
\end{mathletters}

\noindent
in terms of the matrix elements
\begin{mathletters}
\begin{eqnarray}
&&\hat{p}_{n,n'}=2^{-1/2}\left(\sqrt{n+1}\,\delta_{n+1,n'}
 - \sqrt{n}\,\delta_{n-1,n'}\right),\\
&&
\hat{q}_{n,n'}=2^{-1/2}\left(\sqrt{n+1}\,\delta_{n+1,n'}
 + \sqrt{n}\,\delta_{n-1,n'}\right),
\\
&&
\hat{b}_{K,n,n'}=
3^{-1/2}\left[(2n+K+1)\,\delta_{n,n'}
 - \sqrt{(n+1)(n+K+1)}\,\delta_{n+1,n'}
 - \sqrt{n(n+K)}\,\delta_{n-1,n'}\right], \\
&&
\hat{c}_{K,n,n'}=
3^{-1/2}\left[(2n+K+1)\,\delta_{n,n'}
 + \sqrt{(n+1)(n+K+1)}\,\delta_{n+1,n'}
 + \sqrt{n(n+K)}\,\delta_{n-1,n'}\right], \\
&&
\hat{d}_{n,n'}=
\sqrt{(n+1)(n+2)}\,\delta_{n+2,n'}-\sqrt{n(n-1)}\,\delta_{n-2,n'},\\
&&
\hat{e}_{K,n,n'}=
3^{-1/2}\left[\sqrt{(n+1)(n+K+1)}\,\delta_{n+1,n'}
 - \sqrt{n(n+K)}\,\delta_{n-1,n'}\right], \\
&&
\hat{f}^{(+)}_{K,n,n'}=
3^{-1/2}(\sqrt{n+K+1}\,\delta_{n,n'}-\sqrt{n}\,\delta_{n-1,n'}),\\
&&
\hat{f}^{(-)}_{K,n,n'}=
3^{-1/2}(\sqrt{n+1}\,\delta_{n+1,n'}-\sqrt{n+K}\,\delta_{n,n'}),
\\
&&
\hat{g}^{(+)}_{K,n,n'}=
3^{-1/2}(\sqrt{n+K+1}\,\delta_{n,n'}+\sqrt{n}\,\delta_{n-1,n'}),\\
&&
\hat{g}^{(-)}_{K,n,n'}=
3^{-1/2}( \sqrt{n+1}\,\delta_{n+1,n'} +
\sqrt{n+K}\,\delta_{n,n'}),
\\
&&
\hat{z}_{n,n'}=\hat{q}_{n,n'}+\tilde{\omega}^{1/2}_{\rm s}
z_{\rm eq}\delta_{n,n'},
\end{eqnarray}
\label{matrixelements}
\end{mathletters}

\narrowtext

\noindent
with $\gamma=\big( 2\tilde{B}/\tilde{\omega}_{\rm s}\big)^{1/2}$.
These expressions
are straightforward to derive from the standard differential and
recursion relations for the Hermite and generalized Laguerre
polynomials.  It is inconsequential that $C_{K,{\bf n}}=0$,
since Eq.~(\ref{condition}) implies that
$a_{k>0}^{(K,\sigma,{\bf n})}=0$.  Once
Eq.~(\ref{arecursion}) has been solved up to a given order,
the coefficient of the energy expansion corresponding to the
next higher order can be calculated according to
\begin{eqnarray}
&&\epsilon_k
=  \sum_{k'=1}^k\sum_{{\bf n}'}
\Big( H^{(0)}_{k'\!,K,{\bf n},{\bf n}'}
 a_{k-k'}^{(K,\sigma,{\bf n}')}
\nonumber\\
&&\ + H^{(+)}_{k'\!,K,{\bf n},{\bf n}'}
 a_{k-k'}^{(K+1,-\sigma,{\bf n}')}
\! + H^{(-)}_{k'\!,K,{\bf n},{\bf n}'}
 a_{k-k'}^{(K-1,-\sigma,{\bf n}')}\Big) .
\nonumber\\
\label{erecursion}
\end{eqnarray}
In Appendix \ref{app:linalg} we present a convenient linear
algebraic formulation of these recursion relations and an
analysis of the computational cost.

\section{Results}
\label{sec:results}

\subsection{CO$_{\bf 2}$}

An extensive tabulation of rotational spectral lines for
CO$_2$ can be obtained from the HITRAN data base
\cite{hitran}.  We will consider here the ground vibrational
state and the first vibrational ($n_{\rm b}=1$).  Our potential,
Eq.~(\ref{potential}), depends on the 5 parameters
$\omega_{\rm s}$, $\omega_{\rm b}$, $\omega_{\rm a}$, $R_{\rm eq}$,
and $\lambda$, with $z_{\rm eq}=0$.  We will use literature
values for the 3 frequencies \cite{herzberg} and fit
$R_{\rm eq}$ and $\lambda$ to the empirical spectra.

We use an iterative fitting procedure, adjusting $R_{\rm eq}$
to fit the splitting between rotational levels and $\lambda$
to fit the splitting between vibrational levels.  To obtain
initial guesses, we replace $\rho^2$ in the coupling term in
Eq.~(\ref{potential}) with its diagonal matrix element
$(2n_{\rm b}+K+1)/\omega_{\rm b}$, which yields
\begin{equation}
U(R_{\rm s})\approx
{1\over 2}\omega_{\rm s}^2\,
\big(R_{\rm s}-\tilde{R}_{n_{\rm b},K}\big)^2
-{1\over 2}\omega_{\rm s}^2\tilde{R}_{n_{\rm b},K}^2,
\end{equation}
where
\begin{equation}
\tilde{R}_{n_{\rm b},K}=R_{\rm eq}
-{(2n_{\rm b}+K+1)\lambda\over \omega_{\rm s}^2\omega_{\rm b}}.
\end{equation}
Thus, we have independent constants $\tilde{R}_{0,0}$ and
$\tilde{R}_{1,0}$ that can be directly fit to the $n_{\rm b}=0$
and $n_{\rm b}=1$ energies for a given $J$.  These
results yield initial approximations for $R_{\rm eq}$ and
$\lambda$, and hence and initial calculation for the energy.

Let $\tilde{R}_{0,K}^{(1)}$ and $E_{\rm calc}^{(1)}(K)$ be
the initial approximations.  If we assume that for given
$J$ and $K$ the energy is proportional to $\tilde{R}_{0,K}^{-2}$,
then subsequent approximations can be determined using
\begin{equation}
\tilde{R}_{0,K}^{(i+1)}
=\tilde{R}_{0,K}^{(i)}
\left[ E_{\rm calc}^{(i)}(K)/E_{\rm expt}(K)\right]^{1/2},
\end{equation}
for $K=0$ or 1.  For CO$_2$, with $J$ chosen as 84, we found
that 3 iterations were sufficient to obtain convergence.  Our
final paramter values are listed in Table~\ref{table}.

Figures \ref{fig:v0} and \ref{fig:v1} report our results for
the ground and first vibrational states, respectively.
We show the difference between our calculated
energies and energies calculated from a rigid-rotor approximation
using a rotational constant of 0.39 cm$^{-1}$ \cite{herzberg}.
The empirical results \cite{hitran} are indicated by circles.

\subsection{N$_{\bf 2}$O and OCS}

We consider here only the ground vibrational states of the
linear nonsymmetric molecules N$_2$O and OCS.  We set
$\lambda=0$ and fit the bond distances $R_{1,3}$ and
$R_{2,3}$, where the subscript ``3'' labels the central atom
(N or C), and atom 1 is chosen as N for N$_2$O and O for OCS.
According to Eqs.~(\ref{Rcoords}), we have
\begin{equation}
R_{\rm eq}=\left({m_1m_2\over m_{\rm s}}\right)^{1/2}
(R_{1,3}+R_{2,3}),
\end{equation}
\begin{equation}
z_{\rm eq}=\left({m_{\rm s}m_3\over m}\right)^{1/2}
\left[\left({m_1\over m_{\rm s}}\right)R_{1,3}
-\left({m_2\over m_{\rm s}}\right)R_{2,3}\right].
\end{equation}

\noindent
As initial approximations we used the following bond lengths
from the literature \cite{townes}:
$R_{\rm NN}^{(1)}=1.126$ \AA,
$R_{\rm NO}^{(1)}=1.191$ \AA,
$R_{\rm OC}^{(1)}=1.161$ \AA,
$R_{\rm CS}^{(1)}=1.561$ \AA.
To fit the rotational spectrum we simply scaled these
distances according to $R_{ij}=\alpha R_{ij}^{(1)}$ and
obtained the scale factor from a one-parameter fit to a
arbitrarily chosen rotational level.  We obtained
$\alpha_{{\rm N}_2{\rm O}}=1.002$ and
$\alpha_{\rm OCS}=1.0256$.  The resulting values for the
$R_{ij}$, which are listed in Table~\ref{table}, are within
the experimental spread of bond length values.  For example,
our result for $R_{\rm OC}$ agrees with the value
1.22 \AA\ reported by H\"uckel \cite{hueckel}.
Figure~\ref{fig:v0} compares our calculated energies with
empirical \cite{hitran} and rigid-rotor \cite{townes}
results.

\section{Discussion}
\label{sec:conclusions}

The traditional approach to computing molecular spectra is to
numerically diagonalize the Hamiltonian matrix in a finite basis
set.  Our perturbation theory, with an exact recursive
solution at each order, might appear to have little in common
with that method.  Note, however, that our theory also
expresses the wavefunction as a finite linear combination of basis
functions, according to Eq.~(\ref{lincom}).  Diagonalization
uses the variational principle to determine the best set of
coefficients to multiply the basis functions for a given
arbitrary finite truncation of the infinite basis set.  In
contrast, our theory yields the unique asymptotic expansion
of the wavefunction \cite{benderorszag}.
The finite linear combinations of the basis functions are
{\it exact} solutions for the wavefunction
components $\Psi_k$.  Thus, at given order in the
perturbation theory, the asymptotic principle implies a unique
choice of basis set.  Chang {\it et al.} \cite{chang},
incidently, have shown that the basis set chosen by
a perturbation theory is superior to a minimum-energy basis
set for use in numerical diagonalization calculations of
vibrational spectra.

Diagonalization determines the coefficients of the basis
expansion that give the ``best'' value for the energy.
One can expect that the perturbation theory using the same
basis set gives a ``worse'' result \cite{perhaps}.
We have found \cite{suvernev} that it is indeed the
case, for a model vibrational problem, that for a given
basis set perturbation theory yields a less accurate energy
than does diagonalization, and we expect this to be true
as well for the problems considered here.  However, the
computational cost of perturbation theory is much lower.
We show in Appendix~\ref{app:linalg} that
the cost of calculating the perturbation expansion
for a given eigenvalue of a triatomic molecule scales as
$N_v^{5/3}$, where $N_v$ is the size of the vibrational basis.
In contrast, the standard diagonalization algorithm
scales as $N_v^3$ \cite{diag}.

It is of particular interest to consider the effect on cost
of the value of $J$.  For variational calculations
the size of the basis set increases linearly with $J$
\cite{ten}.  Therefore, the cost of diagonalization scales
as $J^3$.  For perturbation theory the total computational
cost scales only linearly with $J$.
Furthermore, the rate of convergence of the perturbation
expansion actually {\it increases} slightly with $J$
\cite{suvernevlet}.  Thus, DPT seems to be an especially
appropriate method for computing eigenvalues corresponding to
high rotational excitation.  Pad\'e summation of the energy
expansion at fifth order yields convergence to 7 decimal
digits for $J=0$ and 8 decimal digits for $J=80$.
We can easily compute these expansions
to 15th order on a desktop computer workstation, obtaining
convergence to more than 12 significant figures.

The relative cost effectiveness of DPT depends on the magnitude
of the accuracy loss, compared to diagonalization, for a given
basis set.  The accuracy of DPT depends on the rate of
convergence of summation approximants of the perturbation
expansion.  In contrast to dimensional expansions for the
electronic Schr\"odinger equation \cite{divergence}, the
expansions for the systems considered here are rapidly
convergent.  A comparison \cite{suvernev} of the cost of
obtaining a given level of accuracy for vibrational
energies of a model 2-coordinate system
of coupled oscillators showed that, in the end, perturbation
theory gave much higher accuracy than diagonalization for a given 
operation count.  We have not carried out diagonalization
calculations for our 3-coordinate systems, so we do not have
a detailed comparison of the methods in this case.  The
accuracy vs. cost of the 3-coordinate perturbation expansion
is comparable to that of the model system, since the rate of
convergence for the model system is slightly slower but
the scaling with $N_v$ is slightly better.  Therefore, we
expect that our method is significantly more efficient than
straightforward diagonalization in this case as well.

The three molecules treated here are tightly bound and therefore
can be accurately described with harmonic, or nearly harmonic,
potentials.  The rate of convergence for potential surfaces
with larger anharmonicity will undoubtedly be slower.
In general one can expect, from fundamental theoretical
considerations \cite{divergence,benderwu}, that dimensional
expansions and coupling-constant expansions for anharmonic
oscillators have a radius of convergence of zero.
Coupling-constant expansions calculated
by \v{C}\'\i\v{z}ek {\it et al.} \cite{cizek}
for model vibrational problems with somewhat larger
anharmonicity did indeed exhibit only semiconvergent behavior,
with divergence setting in at higher orders, although the
expansions were rapidly summable with Pad\'e approximants.
With truly large anharmonicity as in van der Waals molecules
the divergence will undoubtedly be more of a problem.
However, there are techniques that could be used to improve
the rate of convergence in such cases.  For example, one could
redefine the dimensional continuation so that some dependence on
$J$ is present at zeroth order.  This would make it possible to
include centrifugal distortion in the basis functions.  Another
useful technique is to modify the potential energy operator so as
to include an arbitrary renormalization parameter, the value of
which is chosen to optimize the convergence \cite{killingbeck}.
In the context of DPT, this consists of using a
dimensional continuation of the Hamiltonian that has explicit
dimension dependence in the potential, giving the correct
potential at $D=3$ but increasing the stability of the system
at large $D$ \cite{watson}.

In recent years, methods have been developed
\cite{light,tennyson,wyatt,sibert,bowman} that lower the
cost of diagonalization for molecular spectra, especially in
the case of floppy molecules, for which a harmonic-oscillator
basis set converges very slowly.  In view of the
similarities between the form of the wavefunction in
perturbation theory and in diagonalization, it may be possible
to make analogous improvements in the perturbation theory.
In any case, perturbation theory has the advantage
of simplicity and directness, in that a straightforward
recursion relation yields the exact values of the coefficients
of the unique asymptotic expansion.  Furthermore, no
modification of the theory is needed in order to study
resonances, in which case quadratic Pad\'e summation of
the perturbation expansion yields complex energies
\cite{suvernev,resonances}.

Note also that, in
contrast to matrix diagonalization, the perturbation theory
provides each eigenvalue with an unambiguous label,
in terms of the quantum numbers $({\bf n},J,K,\sigma )$.
This labeling is analogous to that provided by the
vibrational SCF method \cite{vscf}; however, that method
introduces a separability approximation while the perturbation
theory, at least in principle, will converge to the exact
result.  In practice, the theory presented here can break
down if the zeroth-order, harmonic, eigenstate is degenerate
or nearly degenerate with another harmonic eigenstate
\cite{crossings}.  Such cases can be treated by
including additional dimension dependence in the
potential energy so as to break the degeneracy
in the large-$D$ limit, although this procedure
results in a somewhat slower rate of convergence.

\acknowledgments{We thank Dr. Martin Dunn for helpful
discussions of hyperdimensional angular momentum.  This work was
supported by the Robert A. Welch Foundation
and the National Science Foundation.}

\appendix

\section{Kinetic Energy Operator in Body-Fixed Coordinates}
\label{app:kineticenergy}

In order to construct the basis functions
$\phi_{J,M,K}(R_{\rm s},\rho,z)$, it is convenient to introduce
Wigner functions \cite{varsh},
\begin{equation}
{\cal D}_{M,M'}^L(\alpha,\beta,\gamma)
=\big\langle L,M\big|
e^{i\alpha \hat{L}_1}e^{i\beta \hat{L}_2}
e^{i\gamma \hat{L}_1}\big| L,M'\big\rangle,
\end{equation}
\begin{equation}
d_{M,M'}^L(\beta)
=e^{-iM\alpha}\,{\cal D}_{M,M'}^L(\alpha,\beta,\gamma)^*\,
 e^{-iM'\gamma},
\end{equation}
where $(\alpha,\beta,\gamma)$ are Euler angles and the $\hat{L}_i$
are components of the orbital angular momentum operator
$\hat{L}$ along perpendicular axes $\hat{\bf x}_i$.
The ${\cal D}$ functions are eigenfunctions of $\hat{L}^2$
and $\hat{L}_i$, and of the component $\hat{L}_{i'}$ along a
body-fixed axis $\hat{\bf x}_{i'}$, with eigenvalues $L(L+1)$,
$M$, and $M'$, respectively.  Let $\phi_{J,M,K}(R_{\rm s},z,\rho)$
be a set of orientation-independent basis functions,
\begin{eqnarray}
&& \Phi({\bf R}_{\rm s},{\bf R}_{\rm a})
\nonumber\\
&&\quad
=\left({2J+1\over 8\pi^2}\right)^{1/2}\sum_{K'}
{\cal D}^J_{M,K'}({\bf \Omega})^*\phi_{J,M,K'}(R_{\rm s},z,\rho).
\end{eqnarray}

\noindent
The kinetic energy operator in this basis is
\begin{equation}
{\cal T}_{J,K,K'}=
{2J+1\over 8\pi^2}\int{\cal D}^J_{M,K}({\bf \Omega})
\,\hat{T}_{\rm rel}\,{\cal D}^J_{M,K'}({\bf \Omega})^*
\,d{\bf \Omega},
\label{TJKKp}
\end{equation}
which is related to $\hat{T}_{\rm rel}$ according to
\begin{eqnarray}
\left({2J+1\over 8\pi^2}\right)^{1/2}
\int && {\cal D}^J_{M,K}({\bf \Omega})\,\hat{T}_{\rm rel}
\,\Phi({\bf R}_{\rm s},{\bf R}_{\rm a})\, d{\bf \Omega}
\nonumber\\
&&\quad
=\sum_{K'}{\cal T}_{J,K,K'}\,
\phi_{J,M,K'}(R_{\rm s},z,\rho ).
\label{Texpansion}
\end{eqnarray}

The wavefunction can also be expanded in terms of a set of
basis functions
$\tilde{\phi}_{J,M,L_{\rm s},L_{\rm a}}(R_{\rm s},R_{\rm a})$,
with quantum numbers $L_{\rm s}$ and $L_{\rm a}$
for the angular momenta corresponding to the vectors
${\bf R}_{\rm s}$ and ${\bf R}_{\rm a}$.  Let ${\bf \Omega}_{\rm s}$
and ${\bf \Omega}_{\rm a}$ be the Euler angles
that describe the orientation of ${\bf R}_{\,\rm s}$ and
${\bf R}_{\,\rm a}$, respectively, relative to the space-fixed
axis.  Then
\widetext
\begin{eqnarray}
\Phi({\bf R}_{\rm s}, {\bf R}_{\rm a})
 =  \frac{1}{4 \pi}\sum_{J,M,L_{\rm s},L_{\rm a}}
&& \sqrt{(2L_{\rm s}+1)(2L_{\rm a}+1)} \;
\tilde{\phi}_{J,M,L_{\rm s},L_{\rm a}}(R_{\rm s}, R_{\rm a})
\nonumber\\
&&\qquad\quad
\times \sum_{M_{\rm s},M_{\rm a}}
C_{L_{\rm s},M_{\rm s};L_{\rm a},M_{\rm a}}^{J, M}
{\cal D}_{M_{\rm s},0}^{L_{\rm s}}({\bf \Omega}_{\rm s})^{\ast} \;
{\cal D}_{M_{\rm a},0}^{L_{\rm a}}({\bf \Omega}_{\rm a})^{\ast}
\label{spheric}
\end{eqnarray}
where the $C_{L_{\rm s}, M_{\rm s}; L_{\rm a}, M_{\rm a}}^{J, M}$
are Clebsch-Gordan coefficients.

The angles that describe the orientation of
${\bf R}_{\,\rm a}$ relative to ${\bf R}_{\,\rm s}$ are
${\bf \Omega}_{\rm a}
-{\bf \Omega}_{\rm s}= (\alpha,\beta,0)$
as illustrated in Fig.~\ref{fig:angles}.  The polar angle
$\beta$ is the angle between ${\bf R}_{\rm a}$ and
${\bf R}_{\rm s}$.  Using the addition theorem for ${\cal D}$
functions and the Clebsch-Gordan series for the product of
${\cal D}$ functions, we find that
\begin{eqnarray}
\Phi({\bf R}_{\rm s}, {\bf R}_{\rm a})
&=& \frac{1}{4 \pi} \sum_{J,M,K}\! \sqrt{2J+1}\,
{\cal D}^J_{M,K}(\alpha_{\rm s},\beta_{\rm s},\alpha)^{\ast}
\nonumber\\
&&\qquad\quad \times
\sum_{L_{\rm s},L_{\rm a}}(-1)^{L_{\rm a}+K}
\sqrt{2L_{\rm a}+1} \;
C^{L_{\rm s}, 0}_{J,-K; L_{\rm a}, K} \;
d^{L_{\rm a}}_{K,0}(\beta)\;
\tilde{\phi}_{J,M,L_{\rm s},L_{\rm a}}(R_{\rm s}, R_{\rm a}).
\end{eqnarray}
Therefore, we can express the former basis functions in terms
of the latter according to
\begin{equation}
\phi_{J,M,K}(R_{\rm s},z,\rho) =
\sum_{L_{\rm s},L_{\rm a}}
(-1)^{L_{\rm a}+K}\sqrt{L_{\rm a}+1/2}
\;C_{J,-K; L_{\rm a}, K}^{L_{\rm s}, 0}
\;d_{K,0}^{L_{\rm a}}(\beta) \;
\tilde{\phi}_{J,M,L_{\rm s},L_{\rm a}}(R_{\rm s}, R_{\rm a}).
\label{basis0}
\end{equation}
On the left-hand side of Eq.~(\ref{basis0}) we have replaced
the internal coordinates $R_{\rm a}$ and $\beta$ with the
cylindrical coordinates defined in Eqs.~(\ref{cylindrical}).

Equation (\ref{basis0}) is useful because the representation
of $\hat{T}_{\rm rel}$ with respect to the
$\tilde{\phi}_{J,M,L_{\rm s},L_{\rm a}}$
is particularly simple.  Suppose we write
\begin{equation}
T_{\rm rel}
= -{1\over 2R_{\rm s}^2}{\partial\over\partial R_{\rm s}}
       R_{\rm s}^2{\partial\over\partial R_{\rm s}}
     +{1\over 2R_{\rm s}^2}\hat{L}^2_{\rm s}
     -{1\over 2R_{\rm a}^2}{\partial\over\partial R_{\rm a}}
       R_{\rm a}^2{\partial\over\partial R_{\rm a}}
     +{1\over 2R_{\rm a}^2}\hat{L}^2_{\rm a}
\label{TLsLa}
\end{equation}

\noindent
and substitute this into Eq.~(\ref{TJKKp}).  Then if we
substitute Eq.~(\ref{basis0}) into Eq.~(\ref{Texpansion}),
we can replace the operator $\hat{L}^2_{\rm s}$ with
$L_{\rm s}(L_{\rm s}+1)$.  It follows that
\begin{eqnarray}
&&\left({2J+1\over 8\pi^2}\right)^{1/2}
\int {\cal D}^J_{M,K}({\bf \Omega})\,\hat{L}^2_{\rm s}
\,\Phi({\bf R}_{\rm s},{\bf R}_{\rm a})d{\bf \Omega}
\nonumber\\
&&\quad
=\,\sum_{L_{\rm s},L_{\rm a}}
\left[{(2L_{\rm s}+1)(2L_{\rm a}+1)\over 2(2J+1)}\right]^{1/2}
 L_{\rm s}(L_{\rm s}+1)\,C^{J,K}_{L_{\rm s},0;L_{\rm a},K}
\, d^{L_{\rm a}}_{K,0}(\beta)\,
\phi_{J,M,L_{\rm s},L_{\rm a}}({\bf R}_{\rm s},{\bf R}_{\rm a}).
\label{Lsexpansion}
\end{eqnarray}

\noindent
This result is in terms of the $\phi_{J,M,L_{\rm s},L_{\rm a}}$,
but these functions can be expressed in terms of the $\phi_{J,M,K}$
by inverting Eq.~(\ref{basis0}),
\begin{eqnarray}
\tilde{\phi}_{J,M,L_{\rm s},L_{\rm a}}(R_{\rm s}, R_{\rm a})
& = & \sqrt{L_{\rm a} + 1/2}
\;\sum_{K'}(-1)^{L_{\rm a}+K'}\;
C_{J,-K'; L_{\rm a}, K'}^{L_{\rm s}, 0}
\nonumber\\
&&\qquad \times
\int_{0}^{\pi}\;
\phi_{J,M,K'}\bigl(R_{\rm s},\;R_{\rm a}\cos \beta ,
\;R_{\rm a} \sin \beta \bigr)\;
d_{K',0}^{L_{\rm a}}(\beta)
\,\sin \beta \, d\beta ,
\label{inverse}
\end{eqnarray}

\noindent
which follows from the orthogonality condition for the
${\cal D}$ functions \cite{edmonds}.

We now evaluate the sum over $L_{\rm s}$ that results from
substituting Eq.~(\ref{inverse}) into Eq.~(\ref{Lsexpansion}).
This substitution gives a product of the Clebsch-Gordan coefficients
can be expressed as an integral over ${\cal D}$ functions,
\begin{eqnarray}
(-1)^{L_{\rm a}}\left({2L_{\rm s}+1\over 2J+1}\right)^{1/2}
&&C^{J,K}_{L_s,0;L_{\rm a},K}
C^{L_{\rm s},0}_{J,-K';L_{\rm a},K'}
=(-1)^K\, C^{L_{\rm s},0}_{J,-K;L_{\rm a},K}
C^{L_{\rm s},0}_{J,-K';L_{\rm a},K'}
\nonumber\\
&&\quad
=(-1)^K\, {2L_{\rm s}+1\over 8\pi^2}
\int {\cal D}^{L_{\rm s}}_{0,0}({\bf \Omega}')^*\,
{\cal D}^J_{-K,-K'}({\bf \Omega}')\,
{\cal D}^{L_{\rm a}}_{K,K'}({\bf \Omega}')\, d{\bf \Omega}'.
\label{CGintegralrep}
\end{eqnarray}
Note that
\begin{equation}
L_{\rm s}(L_{\rm s}+1)\,{\cal D}^{L_{\rm s}}_{0,0}
=\hat{L}_{\rm s}^2\,{\cal D}^{L_{\rm s}}_{0,0}
=\left(\hat{L}_0^2-\hat{L}_{+1}\hat{L}_{-1}
-\hat{L}_{-1}\hat{L}_{+1}\right)\,
{\cal D}^{L_{\rm s}}_{0,0}\, ,
\label{sphericalL}
\end{equation}
where $\hat{L}_0$ and $\hat{L}_{\pm 1}$ are the
components of $\hat{L}_{\rm s}$ in terms of spherical
coordinates.  It follows from Eqs.~(\ref{CGintegralrep}) and
(\ref{sphericalL}), and the orthogonality condition, that
\begin{eqnarray}
&& \sum_{L_{\rm s}}\;
(-1)^{L_{\rm a}+K'}\left({2L_{\rm s}+1\over 2J+1}\right)^{1/2}
 \,L_{\rm s} \,(L_{\rm s}+1)
\,C_{L_{\rm s},0; L_{\rm a},K}^{J,K}
\,C_{J,-K'; L_{\rm a},K'}^{L_{\rm s},0}
=\sum_{L_{\rm s}}\, L_{\rm s}(L_{\rm s}+1)
\,C_{J,-K; L_{\rm a},K}^{L_{\rm s},0}
\,C_{J,-K'; L_{\rm a},K'}^{L_{\rm s},0}
\nonumber\\
&&\qquad\qquad\qquad
 = \big[ J(J+1)-2K^2+L_{\rm a}(L_{\rm a}+1)\big]
\;\delta_{K'\! ,K}
\nonumber\\
&&\qquad\qquad\qquad\quad
-\big\{\big[J(J+1)-K(K-1)\big]
 \big[L_{\rm a}(L_{\rm a}+1)-K(K-1)\big] \big\}^{1/2}
 \;\delta_{K'\! ,K-1}
\nonumber\\
&&\qquad\qquad\qquad\quad
-\big\{\big[J(J+1)-K(K+1)\big]
 \big[L_{\rm a}(L_{\rm a}+1)-K(K+1)\big] \big\}^{1/2}
 \;\delta_{K'\! ,K+1}.
\label{CGidentity}
\end{eqnarray}

\noindent
Thus, the term $\hat{L}^2_{\rm s}/(2R_{\rm s}^2)$ in
$\hat{T}_{\rm rel}$ couples basis functions with neighboring
values of $K$.

Next we need to remove the quantum number $L_{\rm a}$ from
the expression for ${\cal T}_{J,K,K'}$.
We deal with the operator $\hat{L}^2_{\rm a}/(2R_{\rm a}^2)$
by simply replacing it with its differential form,
\begin{eqnarray}
&& \hat{L}^2_{\rm a}\; D_{K,0}^{L_{\rm a}}(\alpha,\beta)^*
\nonumber\\
&&\ = -\left[\frac{\partial^2}{\partial \alpha^2}+\rho^2
\frac{\partial^2}
{\partial z^2} + z^2\left(\frac{1}{\rho}
\frac{\partial}{\partial \rho}
\rho\frac{\partial}{\partial
\rho}+\frac{1}{\rho^2}\frac{\partial^2}{\partial
\alpha^2}\right)-\left(2 z \frac{\partial}{\partial z} + 1
\right)\!\left(\rho
\frac{\partial}{\partial
\rho}+1\right)+1\right] D_{K,0}^{L_{\rm a}}
(\alpha,\beta)^*.
\nonumber\\
\end{eqnarray}
To remove the $L_{\rm a}$ dependence that comes from
the operator $\hat{L}^2_{\rm s}/(2R_{\rm s}^2)$, via
Eq.~(\ref{CGidentity}), we use the fact that
\begin{eqnarray}
\sqrt{L_{\rm a}(L_{\rm a}+1)-K(K\pm 1)}
&& \;D_{K,0}^{L_{\rm a}}(\alpha,\beta)^*
= \pm 2^{1/2}\,\hat{L}_{\mp 1}
 \,D_{K\pm 1,0}^{L_{\rm a}}(\alpha,\beta)^*
\nonumber\\
&&\!\!\!\!
= \pm\,e^{\mp i\,\alpha}
\left(\rho \frac{\partial}{\partial z}
 -z \frac{\partial}{\partial \rho} \pm
 i\,\frac{z}{\rho}\frac{\partial}{\partial\alpha}\right)
\,\left[ e^{i(K\pm 1)\alpha}\,
 d_{K\pm 1,0}^{L_{\rm a}}(\beta)\right]
\label{up_down}
\end{eqnarray}

\noindent
The integration over $\beta$, from
Eq.~(\ref{inverse}), is evaluated using the
completeness property of the $d$ functions:
\begin{equation}
\sum_{L=K}^\infty (L+1/2)\,
d_{\pm K,0}^{L}(\beta) \,d_{\pm K,0}^{L}(\beta')
=\delta(\cos{\beta}-\cos{\beta'}).
\end{equation}
We use the fact that
$i(\partial/\partial\alpha){\cal D}^{L_{\rm a}}_{K,0}
=K{\cal D}^{L_{\rm a}}_{K,0}$ to evaluate the $\alpha$
derivatives.

The two remaining operators in Eq.~(\ref{TLsLa}) are simple
to evaluate in the new basis.  The derivatives with respect to
$R_{\rm s}$ are the same in either basis and the derivatives
with respect to $R_{\rm a}$ are independent
of the orientation of the body-fixed coordinate system.
We can make the substitution
\begin{equation}
-{1\over 2R_{\rm a}^2}{\partial\over\partial R_{\rm a}}
R_{\rm a}^2{\partial\over\partial R_{\rm a}}
+{\hat{L}_{\rm a}^2\over 2R_{\rm a}^2}
=-{1\over 2}{\partial^2\over\partial z^2}
-{1\over 2\rho}{\partial\over\partial\rho}
\rho{\partial\over\partial\rho}
-{1\over 2\rho^2}{\partial^2\over\partial\alpha^2}.
\end{equation}
Eq.~(\ref{relke}) follows after the transformation to
the symmetrized basis functions of Eq.~(\ref{basisfns}).

\section{Linear algebraic formulation of the recursion
relations}

\label{app:linalg}

A linear algebraic method for solving perturbation equations
was recently presented by Dunn {\it et al.} \cite{dunn}.
This approach is completely equivalent to Eqs.~(\ref{arecursion})
and (\ref{erecursion}).  However, it expresses these equations in
a transparent form that simplifies the analysis of
computational cost and is particularly
convenient for efficient implementation on a computer
\cite{dunn,boghosian}.

\subsection{The matrix method}

This method uses the fact that each $\Psi_k$ can be exactly
expressed as a {\it finite} linear combination of the
eigenfunctions of ${\cal H}_0$, according to
Eq.~(\ref{lincom}).  Let ${\bf a}_k^{(K'\! ,\sigma')}$
be the rank-3 tensor of expansion coefficients
$a_k^{(K'\! ,\sigma',{\bf n}')}$, with the tensor
index $(i_1,i_2,i_3)$
corresponding to ${\bf n}'=(i_1-1,i_2-1,i_3-1)$.  Then operators
can be expressed in terms of matrices.  In particular,
Eqs.(\ref{arecursion}) and (\ref{erecursion}) can be written as
\widetext
\begin{equation}
{\bf a}_k^{(K'\! ,\sigma')}=
\widetilde{\bf C}_{K'}\sum_{k'=1}^k
\bigg[ \Big(\epsilon_{k'}{\bf i}\otimes{\bf i}\otimes{\bf i}
 -{\bf H}_{k'\! ,K'}^{(0)}\Big) {\bf a}_{k-k'}^{(K'\!
,\sigma')}
-{\bf H}_{k'\! ,K'}^{(+)}{\bf a}_{k-k'}^{(K'+1,-\sigma')}
-{\bf H}_{k'\! ,K'}^{(-)}{\bf a}_{k-k'}^{(K'-1,-\sigma')}
\bigg],
\label{amateq}
\end{equation}
\begin{equation}
\epsilon_k=\left[{\bf a}_0^{(K,\sigma)}\right]^T\cdot
\sum_{k'=1}^k \Big(
{\bf H}_{k'\! ,K}^{(0)}{\bf a}_{k-k'}^{(K\! ,\sigma)}
+{\bf H}_{k'\! ,K}^{(+)}{\bf a}_{k-k'}^{(K+1,-\sigma)}
+{\bf H}_{k'\! ,K}^{(-)}{\bf a}_{k-k'}^{(K-1,-\sigma)}\Big),
\label{emateq}
\end{equation}
where
\begin{eqnarray}
&&{\bf H}^{(0)}_{k,K} =
(-1)^k {1\over 8}\tilde{\omega}_{\rm s}\gamma^2
\bigg\{
(k+3)\gamma^{k+2}{\bf q}^{k+2}\otimes{\bf i}\otimes{\bf i}
+{4\over 3}\Big[ J(J+1)-(K^2+13/4)\Big](k+1)\gamma^k
  {\bf q}^k\otimes{\bf i}\otimes{\bf i}
\nonumber\\
&&\qquad\qquad
+4k\gamma^{k-1}\bigg[
  \left({\omega_{\rm a}\over\tilde\omega_{\rm b}}\right)
    {\bf q}^{k-1}\otimes{\bf p}^2\otimes{\bf b}_K
 -\left({\tilde\omega_{\rm b}\over\omega_{\rm a}}\right)
    {\bf q}^{k-1}\otimes{\bf z}^2\otimes{\bf c}_K
 + {\bf q}^{k-1}\otimes{\bf d}\otimes{\bf e}_K \bigg] \bigg\}
\nonumber\\
&&\qquad\qquad
\ +\,\delta_{k,1}\,
    {3\lambda
       \over\tilde{\omega}_{\rm s}^{1/2}\tilde{\omega}_{\rm b}}\,
         {\bf q}\otimes{\bf i}\otimes{\bf b}_K,
\\
&&{\bf H}^{(\pm)}_{k,K} =
(-1)^k {1\over 2} \tilde{\omega}_{\rm s}
\sqrt{J(J+1)-K(K\pm 1)}
\nonumber\\
&&\qquad\qquad\qquad\qquad\quad
\times
k \gamma^{k+1}
\bigg[
 \left({\omega_{\rm a}\over\tilde\omega_{\rm b}}\right)^{\! 1/2}
   \!\! {\bf q}^{k-1}\otimes{\bf p}\otimes{\bf f}^{(\pm )}_K
 -\left({\tilde\omega_{\rm b}\over\omega_{\rm a}}\right)^{\! 1/2}
   \!\! {\bf q}^{k-1}\otimes{\bf z}\otimes{\bf g}^{(\pm )}_K
\bigg],
\end{eqnarray}
\begin{equation}
\widetilde{\bf C}_{K'}=\sum_{{\bf n}'}
\big[ (n'_{\rm s}-n_{\rm s})\tilde{\omega}_{\rm s}
   +(n'_{\rm a}-n_{\rm a})\omega_{\rm a}
   +(2n'_{\rm b}+K'-2n_{\rm b}-K)\tilde{\omega}_{\rm b} \big]^{-1}
{\bf i}(n'_{\rm s})\otimes{\bf i}(n'_{\rm a})
\otimes{\bf i}(n'_{\rm b}),
\end{equation}
\noindent
with $\gamma=\big( 2\tilde{B}/\tilde{\omega}_{\rm s}\big)^{1/2}$,
in terms of the matrices $[{\bf i}]_{i,j}=\delta_{i,j}$ and
$[{\bf i}(n)]_{i,j}=\delta_{i,n+1}\delta_{i,j}$, and
the matrices ${\bf p}$, ${\bf q}$, ${\bf b}_K$, ${\bf c}_K$,
${\bf d}$, ${\bf e}_K$, ${\bf f}_K^{(\pm)}$, ${\bf g}_K^{(\pm)}$,
and ${\bf z}$ defined by Eqs.~(\ref{matrixelements}).
We use the direct product notation
${\bf x}_{\rm s}\otimes{\bf x}_{\rm a}
\otimes{\bf x}_{\rm b}\;{\bf a}$
to represent the following procedure:\ multiply the columns of
${\bf a}$, which correspond to the quantum number $n_{\rm b}$,
by the matrix ${\bf x}_{\rm b}$; multiply the west-east rows,
corresponding to $n_{\rm a}$, by ${\bf x}_{\rm a}$; and
then multiply the north-south rows,
corresponding to $n_{\rm s}$, by ${\bf x}_{\rm s}$.
The dot-product notation in Eq.~(\ref{emateq}) represents a
scalar product defined by
\begin{equation}
{\bf a}_k^T\cdot {\bf a}_{k'}
\equiv\sum_{i_1,i_2,i_3}\big[{\bf a}_k\big]_{i_1,i_2,i_3}
\big[{\bf a}_{k'}\big]_{i_1,i_2,i_3}\, .
\end{equation}

The $k$ dependence of the operators ${\bf H}^{(0)}_{k,K}$
and ${\bf H}^{(\pm )}_{k,K}$ is quite simple.  It is found
only in the operations by the coordinate $x$, on account of
the way in which we defined the dimensional continuation of
the kinetic energy operator.  We can take advantage of this
to streamline the form of the recursion relations.
Let us define the operators
\begin{eqnarray}
&&{\bf h}^{(0)}_K=
-{1\over 2}\tilde{\omega}_{\rm s}\gamma^2
\Bigg\{ {1\over 4}\,\gamma^3{\bf q}^3\otimes{\bf i}\otimes{\bf i}
       +{1\over 3}\,\gamma\left[ J(J+1)
                              -\left(K^2+{13\over 4}\right)\right]
         {\bf q}\otimes{\bf i}\otimes{\bf i}
\nonumber\\
&&\qquad\qquad\qquad\qquad
       +\left({\omega_{\rm a}\over\tilde{\omega}_{\rm b}}\right)
         {\bf i}\otimes{\bf p}^2\otimes{\bf b}_K
       -\left({\tilde{\omega}_{\rm b}\over\omega_{\rm a}}\right)
         {\bf i}\otimes{\bf z}^2\otimes{\bf c}_K
       +\,{\bf i}\otimes{\bf d}\otimes{\bf e}_K \Bigg\} ,\\
&&{\bf h}^{(\pm )}_K=
-{1\over 2}\tilde{\omega}_{\rm s}
\sqrt{J(J+1)-K(K\pm 1)}\;\gamma
\left[ \left(
         {\omega_{\rm a}\over\tilde{\omega}_{\rm b}}\right)^{1/2}
         {\bf i}\otimes{\bf p}\otimes{\bf f}^{(\pm )}_K
      -\left(
         {\tilde{\omega}_{\rm b}\over\omega_{\rm a}}\right)^{1/2}
         {\bf i}\otimes{\bf z}\otimes{\bf g}^{(\pm )}_K \right],
\end{eqnarray}
and the tensors
\begin{eqnarray}
&&{\frak a}^{(K,\sigma )}_j=
{\bf h}^{(0)}_K{\bf a}^{(K,\sigma )}_j
+{\bf h}^{(+)}_K{\bf a}^{(K+1,-\sigma )}_j
+{\bf h}^{(-)}_K{\bf a}^{(K-1,-\sigma )}_j,\\
&&{\frak b}^{(K,\sigma )}_j=
-{1\over 8}\tilde{\omega}_{\rm s}\gamma^2
 \left\{ 3\,\gamma^3{\bf q}\otimes{\bf i}\otimes{\bf i}
        +{4\over 3}\gamma\left[ J(J+1)
                              -\left(K^2+{13\over 4}\right)\right]
         {\bf q}\otimes{\bf i}\otimes{\bf i}\right\}
 {\bf a}^{(K,\sigma )}_j. 
\end{eqnarray}

\noindent
Then the recursion relations can be written as
\begin{equation}
{\bf a}^{(K'\! ,\sigma')}_k=\widetilde{\bf C}_{K'}
\Bigg\{\sum_{j=0}^{k-1}
\left[ \epsilon_{k-j}\,{\bf a}^{(K'\! ,\sigma')}_j
+(k-j){\bf A}_{k,j}^{(K'\! ,\sigma')}+{\bf B}_{k,j}^{(K'\! ,\sigma')}
\right]
-{3\lambda\over\tilde{\omega}_{\rm s}^{1/2}\tilde{\omega}_{\rm b}}
  {\bf q}\otimes{\bf i}\otimes{\bf b}_K'\,
  {\bf a}^{(K'\! ,\sigma')}_{k-1}
\Bigg\},
\label{newarecursion}
\end{equation}

\noindent
and
\begin{equation}
\epsilon_k^{(K,\sigma)}=
\left[{\bf a}_0^{(K,\sigma)}\right]^T\!\!\cdot
\Bigg\{
{3\lambda\over\tilde{\omega}_{\rm s}^{1/2}\tilde{\omega}_{\rm b}}
  {\bf q}\otimes{\bf i}\otimes{\bf b}_K\;
  {\bf a}^{(K,\sigma)}_{k-1}
-\sum_{j=0}^{k-1}\left[
(k-j){\bf A}_{k,j}^{(K,\sigma)}+{\bf B}_{k,j}^{(K,\sigma)}\right]
\Bigg\},
\label{newerecursion}
\end{equation}

\narrowtext

\noindent
in terms of the tensors
\begin{mathletters}
\begin{eqnarray}
&&{\bf A}_{k,j}^{(K'\! ,\sigma')}
=\tilde{\bf q}^{k-j-1}\otimes{\bf i}\otimes{\bf i}\;
{\frak a}^{(K'\! ,\sigma')}_j,\\
&&{\bf B}_{k,j}^{(K'\! ,\sigma')}
=\tilde{\bf q}^{k-j-1}\otimes{\bf i}\otimes{\bf i}\;
{\frak b}^{(K'\! ,\sigma')}_j,
\end{eqnarray}
\end{mathletters}

\noindent
where $\tilde{\bf q}=-\gamma{\bf q}$.
Equations (\ref{newarecursion}) and (\ref{newerecursion})
are convenient because the ${\bf A}^{(K'\! ,\sigma')}_{k,j}$ and
${\bf B}^{(K'\! ,\sigma')}_{k,j}$ can be calculated using
recursion relations that are particularly simple:
\begin{mathletters}
\begin{eqnarray}
&&{\bf A}^{(K'\! ,\sigma')}_{k,j}=
\tilde{\bf q}\otimes{\bf i}\otimes{\bf i}\;
{\bf A}^{(K'\! ,\sigma')}_{k-1,j},\\
&&{\bf B}^{(K'\! ,\sigma')}_{k,j}=
\tilde{\bf q}\otimes{\bf i}\otimes{\bf i}\;
{\bf B}^{(K'\! ,\sigma')}_{k-1,j}.
\end{eqnarray}
\label{ABrecursions}
\end{mathletters}

\subsection{Computational cost}

A close examination of Eq.~(\ref{amateq}) reveals
that the columns of ${\bf a}_k^{(K'\! ,\sigma')}$ will have one
more nonzero element than will the columns of
${\bf a}_{k-1}^{(K'\! ,\sigma')}$, the west-east rows will have
two more nonzero elements, and the north-south rows will have
three more nonzero elements.  Therefore, the total number of nonzero
elements in ${\bf a}_k^{(K'\! ,\sigma')}$ scales with order $k$
as $k^3$.  It follows that the number of multiplications
needed at order $k$ in the recursion
to compute $\epsilon_1\,{\bf a}_{k-1}^{(K'\! ,\sigma')}$,
or to compute ${\bf A}_{k,j}^{(K'\! ,\sigma')}$
and ${\bf B}_{k,j}^{(K'\! ,\sigma')}$ using Eqs.~(\ref{ABrecursions}),
scales as $k^3$.

Consider the cost of evaluating
Eq.~(\ref{newarecursion}) for all $k$ such that
$k\le p$, where $p$ is the
order at which the energy expansion will be truncated.
The most expensive part of the calculation is the
evaluation of the 3 terms that are summed over $j$.  For
each value of $j$ there will be a single matrix operation,
scaling at worst as $k^3$, and the sum over $j$ is repeated
for each value of $k$.  Summing the cost over all the values of
$j$ and $k$ gives a cost scaling of $p^5$ for each value of
$(K'\! ,\sigma')$.  The range of $K'$ is $0\le K'\le J$.
Therefore, the total cost scales as $Jp^5$.
According to Eq.~(\ref{lincom}), the number of basis
functions, $N_v$, is equal to the number of nonzero
elements of the corresponding tensors.  Therefore,
$N_v$ scales as $p^3$, and the total
cost scales as $JN_v^{5/3}$.

\vfil\eject

\widetext
\vbox{
\begin{table}
\caption{Frequencies, bond distances, and anharmonicity
constant.}
\begin{tabular}{ccccccc}
molecule&$\omega_{\rm s}$\ \,(cm$^{-1}$)
&$\omega_{\rm a}$\ \, (cm$^{-1}$)&$\omega_{\rm b}$\ \, (cm$^{-1}$)
&$R_{1,3}$ (\AA )&$R_{2,3}$ (\AA )
&$10^{-6}\times\lambda$\ \,(cm$^{-5/2}$)\\
\tableline

CO$_2$&1351.0\protect{\tablenotemark[1]}
&2396.4\protect{\tablenotemark[1]}
&672.2\protect{\tablenotemark[1]}
&1.163647&1.163647&1.79959\\
N$_2$O &1285.0\protect{\tablenotemark[2]}
&2223.5\protect{\tablenotemark[2]}
&588.8\protect{\tablenotemark[2]}
&1.128249&1.193379&0\\
OCS&859.0\protect{\tablenotemark[2]}
&2079.0\protect{\tablenotemark[2]}
&527.0\protect{\tablenotemark[2]}
&1.190693&1.600922&0\\
\end{tabular}
\tablenotemark[1]{Ref. \cite{herzberg}.}
\tablenotemark[2]{Ref. \cite{townes}.}

\label{table}
\end{table} }

\narrowtext

\begin{figure}
\caption{Deviation from rigid-rotor approximation of the
energy levels of the ground vibrational state, vs. the
angular-momentum quantum number $J$.
The ordinate corresponds to $E(J)-E(0)-B_{\rm rot}J(J+1)$,
with $B_{\rm rot}({\rm CO}_2)=0.3906$
\protect{\cite{herzberg}}, $B_{\rm rot}({\rm N}_2{\rm O})=0.419$
\protect{\cite{townes}}, and $B_{\rm rot}({\rm OCS})=0.20286$
\protect{\cite{townes}}.  The solid curves show our
calculated results while the circles show empirical results
\protect{\cite{hitran}}.}
\label{fig:v0}
\end{figure}

\begin{figure}
\caption{Deviation from rigid-rotor approximation of the
energy levels of the first excited vibrational state, vs. the
angular-momentum quantum number $J$.
The ordinate corresponds to $E(J)-E(0)-B_{\rm rot}J(J+1)$,
with $B_{\rm rot}({\rm CO}_2)=0.3906$ \protect{\cite{herzberg}}.
The solid curves show our calculated results while the circles
show empirical results \protect{\cite{hitran}}.}
\label{fig:v1}
\end{figure}

\begin{figure}
\caption{Euler angles $(\alpha_{\rm s},\beta_{\rm s})$ for
the orientation of the body-fixed axis ${\bf R}_{\rm s}$
relative to space-fixed Cartesian axes, and $(\alpha,\beta)$
for the orientation of the vector ${\bf R}_{\rm a}$ relative
to ${\bf R}_{\rm s}$.}
\label{fig:angles}
\end{figure}


\begin{references}
\bibitem{perhaps} It does not necessarily have to be the
case that the energy calculated with perturbation theory will be
less accurate.  Since the perturbation theory is not
variational, it is conceivable that with a sufficiently
optimized dimensional continuation the convergence of the
energy expansion would be so rapid that it would give a
a value for the energy that is more accurate than the
variational result even for the same basis set.
\end{references}
\end{document}